\documentstyle[12pt,aasms4]{article}
\input psfig.sty
\begin{document}

\title{Emissive Mechanism on Spectral Variability of Blazars in High Frequencies}
\author{Jiancheng Wang, Jirong Mao}

\affil{Yunnan Observatory, Chinese Academy of Sciences, P.O. Box 110,
Kunming, Yunnan Province 650011, P. R. China} \affil{National Astronomical
Observatories, Chinese Academy of Sciences} \affil{United Laboratory of
Optical Astronomy, Chinese Academy of Sciences}
\authoremail{jcwang@public.km.yn.cn}

\begin{abstract}
The new results of the evolution of the synchrotron peak for Mrk 421 are
mostly likely related to the particle acceleration process. In order to
account for the above results, we present a model of blazar variability
during the flare in which the emission comes from accelerating electrons. A
diffusion advection equation of the electron energy distribution is derived
to calculate the spectrum and light curve of synchrotron radiation. We
present that the observed shifts of the synchrotron peak moving to higher
energies during the flare are caused by shock acceleration. The observed
relation between changes in the fluxes at specified frequency ranges and
shifts of the peak position is fitted to constrain the physical parameters
of the dissipation region.
\end{abstract}

\keywords{accleration of particles--galaxies: active--
BL Lacertae objects: indivual(Mkn 421)--Plasma:turbulence--radiation
mechanisms: nonthermal--shock waves--X-rays: general}

\section{Introduction}

Among active galactic nuclei, blazars are characterized by a high and
variable degree of polarization and a flux variability often occurring on
very short time scales. The observed continuum is dominated by nonthermal
emission, where the emission deriving from synchrotron and self-Compton is
enhanced by relativistic beaming (Blandford \& Rees 1978). Some blazars
exhibit not only intensity variation extending from radio waves to gamma
rays, but also spectral variation as a function of the flux level in many
frequency bands.

An interpretation of the spectral variations can be envisaged in an
inhomogeneous jet scenario of the kind proposed by Ghisellini et al. (1985)
and Celotti et al. (1991). In this model the structure of the jet is
considered, the magnetic field and the relativistic electron density are
assumed to decrease with distance, and the maximum synchrotron frequency is
assumed to decrease with distance, accounting for a variability time shorter
at larger synchrotron frequencies. The predicted variability pattern is
caused by a disturbance (e.g. shock) traveling down the jet. This
perturbation is assumed to produce only a fixed enhancement in magnetic
field and particle density and to unchange the shape of the electron
distribution. The overall spectrum is the superposition of the located
spectra emitted by each slice of the jet. As it has been applied to BL Lac
objects, the model does not consider the evolution of the relativistic
electron distribution affected by radiation losses, new injection of
relativistic electrons and shock waves. However, simultaneous observations
of blazars over a wide spectral range suggest that significant radiation
losses and injection of relativistic electrons occur in flares. For example,
the variability of PKS2155-304 at X-ray, EUV and UV bands is decreased in
amplitude and with significant delays approximately satisfying a $\nu
^{-\frac 12}$ relation (Urry et al. 1997). In particular, the radiation loss
time for electrons that emit optical and X-ray emission in blazars is
probably far less than the light (or the shock) crossing time of the
emitting region. Consequently, if the electron distributions responsible for
high energy emission are kept by the shock acceleration processes, then
radiation losses and the injection of electrons occur in the vicinity of the
shock during the flare and must be considered in the theory. Many models to
reproduce the spectral variability have been developed ( Mastichiadis \&
Kirk 1997; Georganopoulous \& Marscher 1998; Kirk, Rieger \& Mastichiadis
1998; Wang et al. 1999; Ghiaberge \& Ghisellini 1999; Li \& Kusunose 2000).
The observed spectrum and time delays between the light curves at fluxes at
different frequencies are believed to be produced by the electron
distribution at different stages of evolution after episodic electron
injection phases. However these models can not account for a synchrotron
peak drifting to higher energies at X-ray wavelengths during the rising
phase of the flare in Mrk 421 (Fossati et al. 2000). The above hard lag is
most likely related to the particle acceleration process. The role of
particle acceleration by shock waves has been considered by Kirk et al.
(1998). In this model, electron acceleration and radiation zones are
separated. Electrons are continuously accelerated at the shock front, and
subsequently drift away from it in the downstream and emit most of the
radiation. The emission from the acceleration zone is ignored.

Through shocks can quickly accelerate particles to very high energies, this
requires the existence of some scattering agent to force repeated passage of
the particles across the shock. The most likely agent for scattering is
plasma turbulence or plasma waves. However, the plasma turbulence needed for
the scattering can not only accelerate particles stochastically (second
order Fermi acceleration), but can cause particle diffusion in the
downstream zone quickly. If the dissipation of the bulk energy of blazar jet
into particles includes shocks and plasma turbulence, it seems difficult to
distinguish acceleration zone and emission region. The emitting particles in
downstream could be re-accelerated by shock waves under the scattering of
plasma turbulence. Therefore we try to consider the synchrotron emission
from a single dissipation region which includes shock and turbulence
processes and radiation losses.

In Section II we present a stochastic differential equation to describe a
particle acceleration and cooling processes, together with the assumptions
made. Then we derive a diffusion-advection equation of particle energy
distribution which simulates the temporal evolution of the particle
radiation occurring when shock waves and plasma turbulence are produced in a
dissipation region of a jet. In section III we apply our model to Mrk 421 to
explain the hard lag observed in the X-ray light curves. Finally, we draw
our conclusions in section IV.

\section{ The Model}

We focus on the particle acceleration and cooling processes and assume that
pitch-angle scattering maintains an almost isotropic particle distribution.
We also assume that the shock in a relativistic blazar jet is
nonrelativistic, and assume that the light crossing time of the dissipation
region is short compared with the intrinsic cooling and acceleration
timescales. It is shown that the observed variability is determined by the
intrinsic cooling and acceleration processes. We use a homogeneous model in
which both the magnetic field and the particle distribution function are
assumed homogeneous through the dissipation region, and consider only the
synchrotron radiation of the accelerating particles, leaving the more
involved computation of inverse Compton emission to future work.

\subsection{Evolution of particle energy distribution}

Firstly we use a stochastic differential equation (SDE) to describe a
particle acceleration and cooling processes, as has been used by some
authors (Kr\"ulls \& Achterberg 1994; Mastichiadis \& Kirk 1997). Then we
derive a diffusion-advection equation of particle energy distribution (or
Fokker-Plank equation). Because the particle distribution is assumed to be
isotropic in space and momentum, we consider only the evolution of the
particle energy given by

\begin{equation}
\frac{d\varepsilon }{dt}=\stackrel{\cdot }{\varepsilon }_{gain}+\stackrel{%
\cdot }{\varepsilon }_{loss}.
\end{equation}
The energy losses $\stackrel{\cdot }{\varepsilon }_{loss}$ is mainly
determined by synchrotron radiation and is a deterministic process in
homogeneous model. It has the form

\begin{equation}
\stackrel{\cdot }{\varepsilon }_{loss}=-\beta \varepsilon ^2,\qquad \beta =%
\frac{32\pi e^4}{9m_e^4c^3}\left( \frac{B^2}{8\pi }\right) .
\end{equation}
The energy gains $\stackrel{\cdot }{\varepsilon }_{gain}$ is due to shock
wave and plasma turbulence, and is assumed to have the form

\begin{equation}
\stackrel{\cdot }{\varepsilon }_{gain}=a\varepsilon ,
\end{equation}
where $a$ is a stochastic variable which is determined by mean value $%
\left\langle a\right\rangle =\alpha $ and a variable $\zeta (t)$ given by
the Gaussian noise process, namely it defined by the value $\left\langle
\zeta (t)\right\rangle =0$ and the autocorrelation function $\left\langle
\zeta (t)\zeta (t^{\prime })\right\rangle =2D\delta (t-t^{\prime })$, where
the coefficient $D$ denotes the noise intensity which is assumed to
represent the stochastic influence of plasma turbulence. This means that the
turbulence will influence the rate of shock acceleration stochastically and
cause the particle energy gain to be a random process. To that end one
writes Eq(2) as an SDE

\begin{equation}
\frac{d\varepsilon }{dt}=\alpha \varepsilon -\beta \varepsilon
^2+\varepsilon \zeta (t).
\end{equation}
For simplify we have assumed that $\alpha $, $\beta $ and $D$ change slowly
compared to particle cooling and acceleration processes, and treat them as
constants.

The Fokker-Plank equation corresponding to the system of SDE given in Eq(4)
are (Stratonovich 1967)

\begin{equation}
\frac{\partial f(\varepsilon ,t)}{\partial t}=-\frac \partial {\partial
\varepsilon }\left[ (\alpha \varepsilon -\beta \varepsilon ^2+D\varepsilon
)f(\varepsilon ,t)\right] +D\frac{\partial ^2}{\partial ^2\varepsilon }%
\left[ \varepsilon ^2f(\varepsilon ,t)\right] ,
\end{equation}
where $f(\varepsilon ,t)$ denotes a conditional probability density $%
p(\varepsilon ,t/\varepsilon _0,t_0)$ for any initial energy $\varepsilon _0$
at $t_0$. If the number $N_0$ of particles undergo acceleration, the
evolution equation of the particle number density $N(\varepsilon
,t)=N_0f(\varepsilon ,t)$ is similar to Eq(5). In order to include particle
escape and injection, we are easy to extend the evolution equation as
(Schlickeiser 1984)

\begin{equation}
\frac{\partial N(\varepsilon ,t)}{\partial t}=-\frac \partial {\partial
\varepsilon }\left[ (\alpha \varepsilon -\beta \varepsilon ^2)N(\varepsilon
,t)\right] +D\frac{\partial ^2}{\partial ^2\varepsilon }\left[ \varepsilon
^2N(\varepsilon ,t)\right] -\frac{N(\varepsilon ,t)}{T_{esc}}+Q(\varepsilon
,t).
\end{equation}
We consider only the intrinsic cooling and acceleration processes and assume
that the timescale of particle is longer than that of the intrinsic physical
processes. We ignore the particle escape term in Eq(6). Otherwise we assume
that the particles in the dissipation region are accelerated from the
initial time $t_0$ and no new particle is injected in the following time.
The injection term in Eq(6) is also ignored. We scale the energy $%
\varepsilon $ and the time $t$ as $\varepsilon =\varepsilon _lX$ and $%
t=\alpha ^{-1}\tau $, where the critical energy $\varepsilon _l=\alpha \beta
^{-1}$, and also define $\kappa =D\alpha ^{-1}$ which is the ratio of shock
acceleration timescale to turbulent acceleration timescale. We can now
rewrite the evolution equation of the particle distribution as

\begin{equation}
\frac{\partial N(X,\tau )}{\partial \tau }=-\frac \partial {\partial
X}\left[ X-X^2+\delta X\right] N(X,\tau )+\kappa \frac{\partial ^2}{\partial
^2X}\left[ X^2N(X,\tau )\right] .
\end{equation}

We now turn to finding the solution of Eq(7) describing the evolution of
particle distribution. Assume the initial energy distribution of the
particles to be $N(X,0)=\delta (X-x_0)$. We firstly consider the solution of
Eq(7) in the case of weak interaction of plasma turbulence $\kappa \ll 1$.

We write $X$ as $X=x(\tau )+\sqrt{\kappa }y$, where $x(\tau )=\left\langle
X\right\rangle =\int XN(X,\tau )dX$. Multiplying Eq(7) by $X$ and $X^2$ and
integrating $X$, we obtain the equation of moments

\begin{equation}
\frac{dx(\tau )}{d\tau }=x(\tau )-x^2(\tau ),
\end{equation}

\begin{equation}
\frac{d\sigma (\tau )}{d\tau }=2\left[ 1-x(\tau )\right] \sigma (\tau
)+2x^2(\tau ),
\end{equation}
where $\sigma (\tau )=\left\langle y^2\right\rangle =\int y^2N(X,\tau )dX$.
Other high order moments can be ignored due to $\kappa \ll 1$. Therefore,
the solution of Eq(7) is approximately given by a Gaussian distribution

\begin{equation}
N(X,\tau )=\frac 1{\sqrt{2\pi \delta \sigma (\tau )}}\exp \left\{ -\frac{%
\left[ X-x(\tau )\right] ^2}{2\delta \sigma (\tau )}\right\} ,
\end{equation}
where $x(\tau )$ and $\sigma (\tau )$ are respectively the solutions of
Eq(8) and Eq(9) which are given by

\begin{equation}
x(\tau )=\left[ 1+(x_0^{-1}-1)e^{-\tau }\right] ^{-1},
\end{equation}
and

\begin{equation}
\sigma (\tau )=C\left[ x(\tau )\right] -C(x_0).
\end{equation}
The function $C(x)$ is defined as

\begin{equation}
C(x)=2(x-x^2)^2\ln \left( \frac x{1-x}\right) +x^4+4x^3(1-x)
\end{equation}
shown in Fig.1. Clearly the energy distribution is narrow, centered at a
typical energy $\varepsilon _lx(\tau )$ and with width $\varepsilon _l\sqrt{%
\kappa \sigma (\tau )}$, quickly evolving with time to $\varepsilon _l$ and $%
\varepsilon _l\sqrt{\kappa }$, where $\varepsilon _l$ corresponds to the
highest energies. It is shown that the particle acceleration is limited by
radiation losses

We now consider the solution of Eq(7) for the general value of $\kappa $. It
is noted that Eq(7) has a stationary solution given by

\begin{equation}
N_c(X)=CX^{\frac 1\kappa -1}\exp \left( -\frac X\kappa \right) ,
\end{equation}
where $C$ is a normalization constant. The function $N_c(X)$ has a peak at $%
X=1-\kappa $ and a narrow shape with decreasing $\kappa $ which is shown in
Figure 2. It indicates that the turbulence leads the spread of the particle
energy distribution. The formation of a peak is caused by the interplay of
shock wave acceleration, turbulence acceleration and radiation losses

The time-dependent solution can be constructed by eigenfunctions and
eigenvalues of Eq(7) (Schenzle \& Brand 1979):

\begin{equation}
N(X,\tau )=\sum_\mu e^{-\lambda _\mu \tau }C_{\lambda _\mu }P_{\lambda _\mu
}(X),
\end{equation}
where $P_{\lambda _\mu }(X)$ denotes the eigenfuctions of the operator $%
\widehat{A}(X)$ with eigenvalues $\lambda _\mu $, $C_{\lambda _\mu }$ are
expansion coefficients,

\begin{equation}
\widehat{A}(X)P_{\lambda _\mu }(X)\equiv -\frac \partial {\partial X}\left[
(X-X^2+\delta X)P_{\lambda _\mu }(X)\right] +\kappa \frac{\partial ^2}{%
\partial X^2}\left[ X^2P_{\lambda _\mu }(X)\right] =\lambda _\mu P_{\lambda
_\mu }(X),
\end{equation}
where $\widehat{A}(X)$ is not an Hermtean operator. The condition of
square-integrability and the correlation for the eigenfunctions are

\begin{equation}
\int_0^\infty N_s(X)\left[ P_{\lambda _\mu }(X)\right] ^2dX=1,\qquad
\int_0^\infty \overline{P}_{\lambda _\mu ^{\prime }}(X)P_{\lambda _\mu
}(X)dX=\delta (\lambda _\mu ^{\prime }-\lambda _\mu ),
\end{equation}
where the function $\overline{P}_{\lambda _\mu ^{\prime }}(X)=P_{\lambda
_\mu }(X)/N_s(X)$. From the above condition, the allowed eigenvalues and
eigenfunctions are given by

\begin{equation}
\lambda _n=n-\kappa n^2,
\end{equation}
\begin{equation}
P_{\lambda _n}(X)=N_s(X)X^{-n}M(-n,-2n+1+\kappa ^{-1},X\kappa ^{-1})
\end{equation}
for the discrete part of the eigenvalue spectrum, where $n$ is the positive
integer subject to the condition $n<\frac 12\kappa ^{-1}$, and a continuous
spectrum for $\lambda _\mu >\frac 1{4\kappa }$ with the corresponding
eigenfunctions

\begin{equation}
P_{\lambda _\mu }(X)=N_s(X)X^{i\sigma -\frac 1{2\kappa }}U(i\sigma -\frac
1{2\kappa },2i\sigma +1,X\kappa ^{-1}),
\end{equation}
where $M(a,b,X)$ and $U(a,b,X)$ are Kummer's function (Abramwitz \& Stegun
1970) and $\sigma =\left( \lambda _\mu \delta \kappa -\frac 14\kappa
^{-2}\right) ^{1/2}$. The expansion coefficients $C_{\lambda _\mu }$ are
determined from the initial condition

\begin{equation}
C_{\lambda _\mu }=\int_0^\infty N_s^{-1}(X)P_{\lambda _\mu }(X)N(X,0)dX.
\end{equation}
One finally finds

\begin{equation}
N(X,\tau )=\sum_ne^{-\lambda _n\tau }C_{\lambda _n}P_{\lambda
_n}(X)+\int_{\frac 1{4\delta ^2}}^\infty e^{-\lambda _\mu \tau }C_{\lambda
_\mu }P_{\lambda _\mu }(X)d\lambda _\mu .
\end{equation}

We now turn to consider the evolution of the particle distribution when the
particle acceleration stops and ignore the particle escape. The evolution
equation of the particle distribution is given by

\begin{equation}
\frac{\partial N(\varepsilon ,t)}{\partial t}=\frac \partial {\partial
\varepsilon }\left[ \beta \varepsilon ^2N(\varepsilon ,t)\right]
+N_s(\varepsilon )\delta (t-t_s),
\end{equation}
where $N_s(\varepsilon )$ and $t_s$ are respectively the particle
distribution and the time when the acceleration stops. This case corresponds
to the decay phase. The solution of Eq(23) is

\begin{equation}
N(\varepsilon ,t)=N_s(\varepsilon b^{-1})b^{-2},\qquad b=\left[ 1-\beta
\varepsilon (t-t_s)\right] \geq 0
\end{equation}
which is an extension of the solution given by Kazanas et al. (1998).

\section{Application to Mrk 421}

A peaked spectrum from observations constrains particle energy distribution
to be narrow. It indicates that shock acceleration will dominate particle
acceleration processes corresponding to $\kappa \ll 1$. In the following
text we focus on the calculation of light curves and spectra in the case of $%
\kappa \ll 1$. We assume that the initial energy distribution of electrons
to be single energy distribution with initial energy $x_0\varepsilon _l$.
The synchrotron emission is calculated from the energy distribution of
electrons which is parameterized by the critical energy $\varepsilon _l$. We
first estimate the parameter $\varepsilon _l$ from the timescales of shock
acceleration and radiation losses. The timescale $\alpha ^{-1}$ of shock
acceleration can be estimated in terms of the light crossing timescale $%
\Delta t$. We assume the timescale $\alpha ^{-1}$ to be $\zeta $ times the
light crossing timescale, e.g., $\alpha ^{-1}=\zeta \Delta t$, one obtain

\begin{equation}
\frac{\varepsilon _l}{m_ec^2}=8.95\times 10^5\left( \frac{\zeta \Delta t}{day%
}\right) ^{-1}\left( \frac B{0.1G}\right) ^{-2}.
\end{equation}
The critical frequency is given by

\begin{equation}
\nu _l=5.82\left( \frac{\zeta \Delta t}{day}\right) ^{-2}\left( \frac
B{0.1G}\right) ^{-3}keV.
\end{equation}

We now calculate the synchrotron emissivity as a function of time and
frequency $F_\nu (\nu ,t)$ with

\begin{equation}
F_\nu (\nu ,t)\propto \int d\varepsilon N(\varepsilon ,t)P(\nu ,\varepsilon
).
\end{equation}
$P(\nu ,\varepsilon )$ is the single particle synchrotron emissivity which
has a approximate function given by Kaplan \& Tsytovich (1973):

\begin{equation}
P(\nu ,\varepsilon )=a_s\sqrt{3}\left( \frac 32\right) ^{1/3}F\left( \frac
\nu {\nu _c}\right) ,
\end{equation}
where $a_s=2\pi \sqrt{3}e^2c^{-1}\nu _L\sin \theta $ is a constant and $\nu
_c=3\varepsilon ^2\nu _L\sin \theta /(2m_e^2c^4)$, with $\nu _L=eB/(2\pi
m_ec)$ the electron Lamor frequency and $\theta $ the angle between the
magnetic field direction and the line of sight. The function $F(X)$ is $%
F(X)=X^{1/3}\exp (-X)$.

Because the particle distribution is a narrow Guassian distribution, it can
be approximated as a Delta function, e.g., $N(\varepsilon ,t)=N_0\delta
(\varepsilon -\varepsilon _c)$, where $\varepsilon _c$ is the central energy
of Guassian distribution which is evolving with time ($\varepsilon _c=x(\tau
)\varepsilon _l$). The synchrotron emission is approximated as $F_\nu
\propto P(\nu ,\varepsilon _c)$. In Figure 3 we plot the synchrotron
spectrum $\nu F_\nu $ of accelerating electrons at different times. Clearly
the synchrotron peak shifts to higher frequencies during the rise. It
indicates that the presence of a hard lag is caused by the emission from the
electrons which are being accelerated to higher energies by shock wave. The
highest energy is determined by the acceleration and cooling rates.

When the shock acceleration stops, the particle distribution will evolve
according to Eq(24) due to radiation losses. The corresponding synchrotron
emission is given by

\begin{equation}
F_\nu \propto \int_0^{\varepsilon ^{*}}N_s(\varepsilon b^{-1})b^{-2}P(\nu
,\varepsilon )d\varepsilon ,
\end{equation}
where $\varepsilon ^{*}$ is a cut-off frequency which comes from the limit
of $b=\left[ 1-\beta \varepsilon (t-t_d)\right] \geq 0$, e.g., $\varepsilon
^{*}=\beta ^{-1}(t-t_s)^{-1}$. Using a Delta function to approximate $N_d$,
e.g., $N_d(\varepsilon b^{-1})=N_0\delta (\varepsilon b^{-1}-\varepsilon _c)$%
, we obtain the synchrotron emission as

\begin{equation}
F_\nu \propto \int_0^{\varepsilon ^{*}}\delta (\varepsilon
b^{-1}-\varepsilon _c)b^{-2}P(\nu ,\varepsilon )d\varepsilon =P(\nu
,\varepsilon _t),
\end{equation}
where $\varepsilon _t=\varepsilon _c[1+\varepsilon _c\beta (t-t_s)]^{-1}$
and $\varepsilon _c$ is the central energy of particle Guassian distribution
when the shock acceleration stops. Clearly the peak of $\nu F_\nu $
decreases to lower frequencies as soon as the shock is over.

The brightness of synchrotron emission at specified frequency ranges $(\nu
_1,\nu _2)$ for the flare is simply given by

\begin{equation}
L_f=\int F_\nu d\nu \propto \int P(\nu ,\varepsilon _c)d\nu .
\end{equation}
The peak frequency of synchrotron spectrum $\nu F_\nu $ locates at $\nu
_p=\frac 43\nu _c$. The relation of $\nu _p$ and $L_f$ is

\begin{equation}
L_f\propto Q(\nu _p,\nu _{1,}\nu _2)=\nu _p\int_{3\nu _1/4\nu _p}^{3\nu
_2/4\nu _p}X^{\frac 13}e^{-X}dX,
\end{equation}
where the function $Q(\nu _p,\nu _{1,}\nu _2)$ is shown in Figure 4. The
peak frequencies show an obviously linear relation with the fluxes in
0.1-10keV ranges. This relation reveals the formation of a narrow particle
energy distribution by shock wave and the subsequent evolution by radiation
losses.

In order to show the above results for application, we thus study the
relation between changes in the brightness and shifts of the peak position
during the flare based on data of Mrk 421 on 1998 April 21 (Fossati et al.
2000). We introduce a steady spectrum $b(\nu )$ which does not take part in
the flare. The observed brightness $L_{obs}$ will include two contributions
of the flaring and steady components given by

\begin{equation}
L_{obs}=A\left[ Q(\nu _p,\nu _{1,}\nu _2)+K(\nu _1,\nu _2)\right] ,
\end{equation}
where $K(\nu _1,\nu _2)=\int_{\nu _1}^{\nu _2}b(\nu )d\nu $, and $A=a_s\sqrt{%
3}\left( \frac 32\right) ^{1/3}n_0\left( \frac{c\Delta t}{D_L}\right)
^2c\Delta t$ which is given by

\begin{equation}
A=3.0\times 10^{-10}\left( \frac{n_0}{10^7cm^{-3}}\right) \left( \frac
B{0.1G}\right) \left( \frac{\Delta t}{hr}\right) ^3ergs^{-1}cm^{-2}keV^{-1},
\end{equation}
where $n_0$ is the number density of accelerated particles. $B$ and $R$ are
the magnetic field and size of dissipation region respectively. $D_L$ is the
luminosity distant of source to the observer. We take the Hubble constant to
be $50kms^{-1}Mpc^{-1}$and the redshift of Mrk 421 to be $0.031$. The fits
of the model are shown in Figure 5. The values of the model parameters are $%
A=10.0\times 10^{-10}ergs^{-1}cm^{-2}keV^{-1}$and $K=0.08,0.23,0.41keV$ for
the 0.2-1, 2-10, 0.1-10keV fluxes. Thus the relation between the spectral
evolution and the flux variability during the flare can constrain the
physical parameters of dissipation regions. It is important to notice that
the inclusion of a steady component is crucial to fit the above relations.
The fits of the model indeed achieve the deconvolution of the spectral
energy distribution into different contributions. The changes of $K(\nu
_1,\nu _2)$ with different frequency ranges show that the steady emission
concentrates at 2-10keV ranges.

The evolution of synchrotron emission with times is given by Eq(30), where
the central energy $\varepsilon _t$ evolves during the acceleration and
post-acceleration phase as

\begin{equation}
\varepsilon _t=\left\{ 
\begin{array}{cc}
\varepsilon _l[1+(x_0^{-1}-1)e^{-\tau }]^{-1} & t\leq t_s \\ 
\varepsilon _t=\varepsilon _ly(\tau _s)[1+y(\tau _s)(\tau -\tau _s)]^{-1} & 
t>t_s
\end{array}
\right. ,
\end{equation}
where $y(\tau _s)=[1+(x_0^{-1}-1)e^{-\tau }]^{-1}$, $\tau =\alpha t=\frac
t{\zeta \Delta t}$ and $\tau _s=\frac{t_s}{\zeta \Delta t}$. Figure 6 shows
the function $F(\nu /\nu _c)$ of light curves in different frequencies. We
find the following interesting results. During the acceleration, the higher
energy emissions lag the lower energy ones. The light curve is approximately
symmetric in the lower energy bands, and it becomes increasingly asymmetric
at higher energies. During the post-acceleration, the light curves are
traced in the opposite way. The higher energy emissions first rise and then
decays rapidly. The light curve is symmetric in various energies due to the
single rise and decay timescale determined by cooling timescale. It should
be also noted that the emissions in frequency ranges of $\frac 13\nu
_lx_0^2<\nu <\frac 13\nu _l$ have the light curves of the peak shape due to
the maximum value of function $F(\nu /\nu _c)$ occurring at $\nu =\frac
13\nu _l[1+(x_0^{-1}-1)e^{-\tau }]$. The light curves in $\nu \geq \nu _l$
bands successively increase to the maximum value of $F(\nu /\nu _c)$ until
the shock wave acceleration stops and then rapidly decay due to particle
radiation losses. If the particle escape is not ignored, the variability
amplitude in lower energies will decreases in the post-acceleration (Wang et
al. 1999). If there is new particle injection during the shock acceleration,
higher energy particles will be more than lower energy ones. The larger
amplitude variability appears at higher energies.

\section{Conclusions}

The recent BeppoSAX observations of Mrk 421 (Fossati et al. 2000) provide
important information to understand particle acceleration processes. Within
a single emission region scenario for blazar jets, we have studied the time
dependent behavior of the particle distribution affected by the particle
acceleration and radiation losses, and calculated the form of light curves
and spectra at different times. We have presented that the observed shifts
of the synchrotron peak moving to higher energy during the flare are caused
by shock acceleration. The accelerating particles follow a narrow Guassian
distribution with a central energy quickly evolving with time to the highest
energy when the shock acceleration dominates the turbulent acceleration. The
highest energy is limited by radiation losses. Our results are important for
the observed fast dissipation region of blazar jets where the light crossing
time of the region is shorter than the particle acceleration and cooling
time scales. The observed fast variability indicates the particle
acceleration and cooling processes.

The observed relation between changes in the fluxes at specified frequency
ranges and shifts of the peak position during the flare can estimate the
magnetic field and the number density of accelerated particles in the
dissipation region.

An energy dependence of the shape of the light curve observed in Mrk 421
during the flare (Fossati et al. 2000) can be connected to the particle
acceleration and cooling time scales. During the acceleration, the observed
light curves are expected to be symmetric at low energies where the
acceleration time is similar to the cooling time and much longer than the
light crossing time. The asymmetric light curves (faster rise) occur at
higher energies where the acceleration time is shorter than the cooling time
and is comparable to the light crossing time. During the post-acceleration,
the light curve is symmetric where the rise and decay timescales are
equivalent to cooling time scale.

With the steady state solution of the diffusion-advection equation of
particle energy distribution, we have demonstrated that the turbulent
acceleration leads the particle energy diffusion. The strength of turbulent
acceleration determines the width of energy diffusion which modifies the
particle distribution with respect to the pure shock acceleration case.

\acknowledgements
We acknowledge the support of {National Astronomical Observatories} grants
(99-5102CA), Chinese Academy of Sciences (CAS).

\clearpage
\begin{figure}
\centerline{\psfig{figure=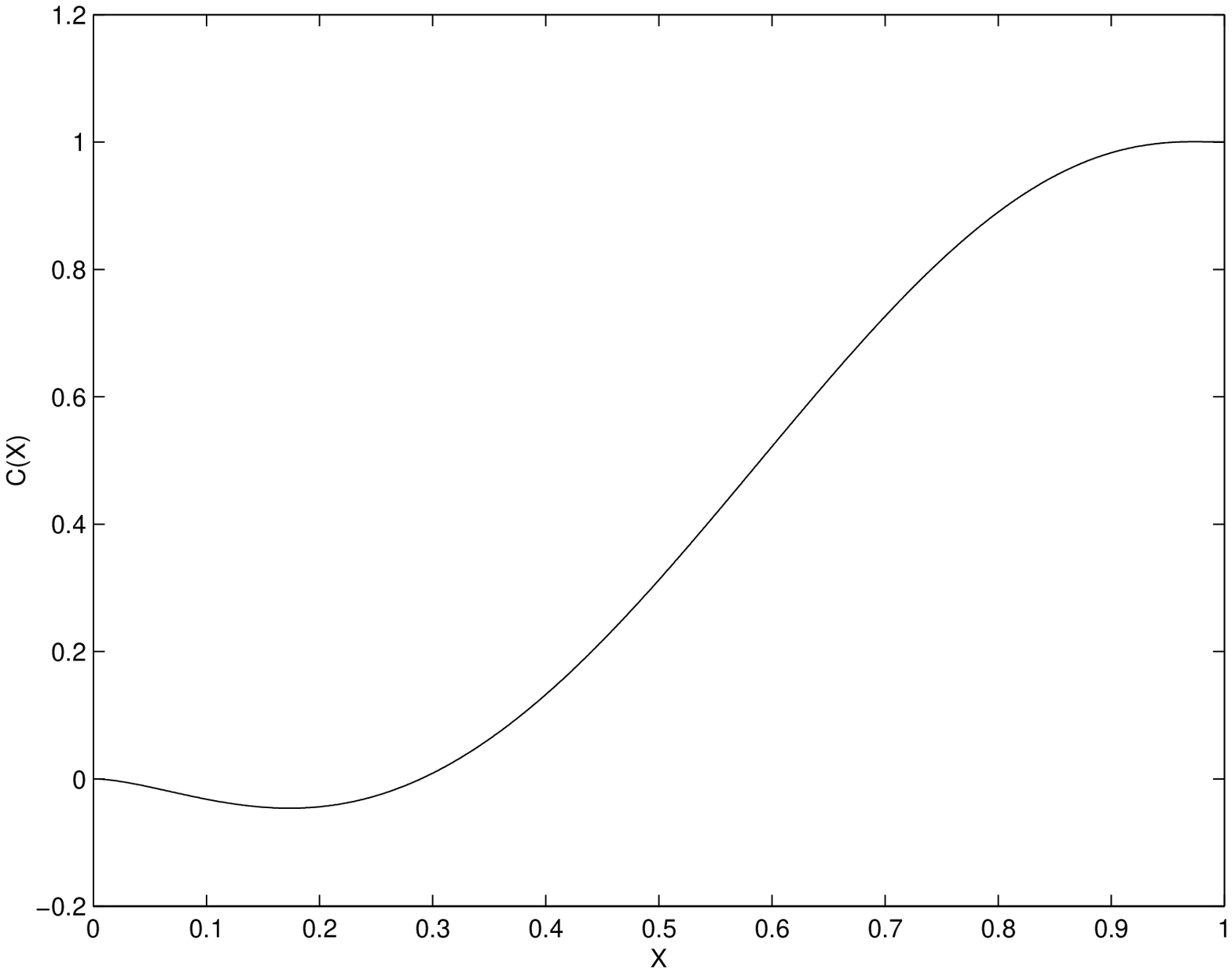}}
\caption{Evolution of the squared variance of particle Guassian energy
distribution}
\label{fig1}
\end{figure}
\begin{figure}
\centerline{\psfig{figure=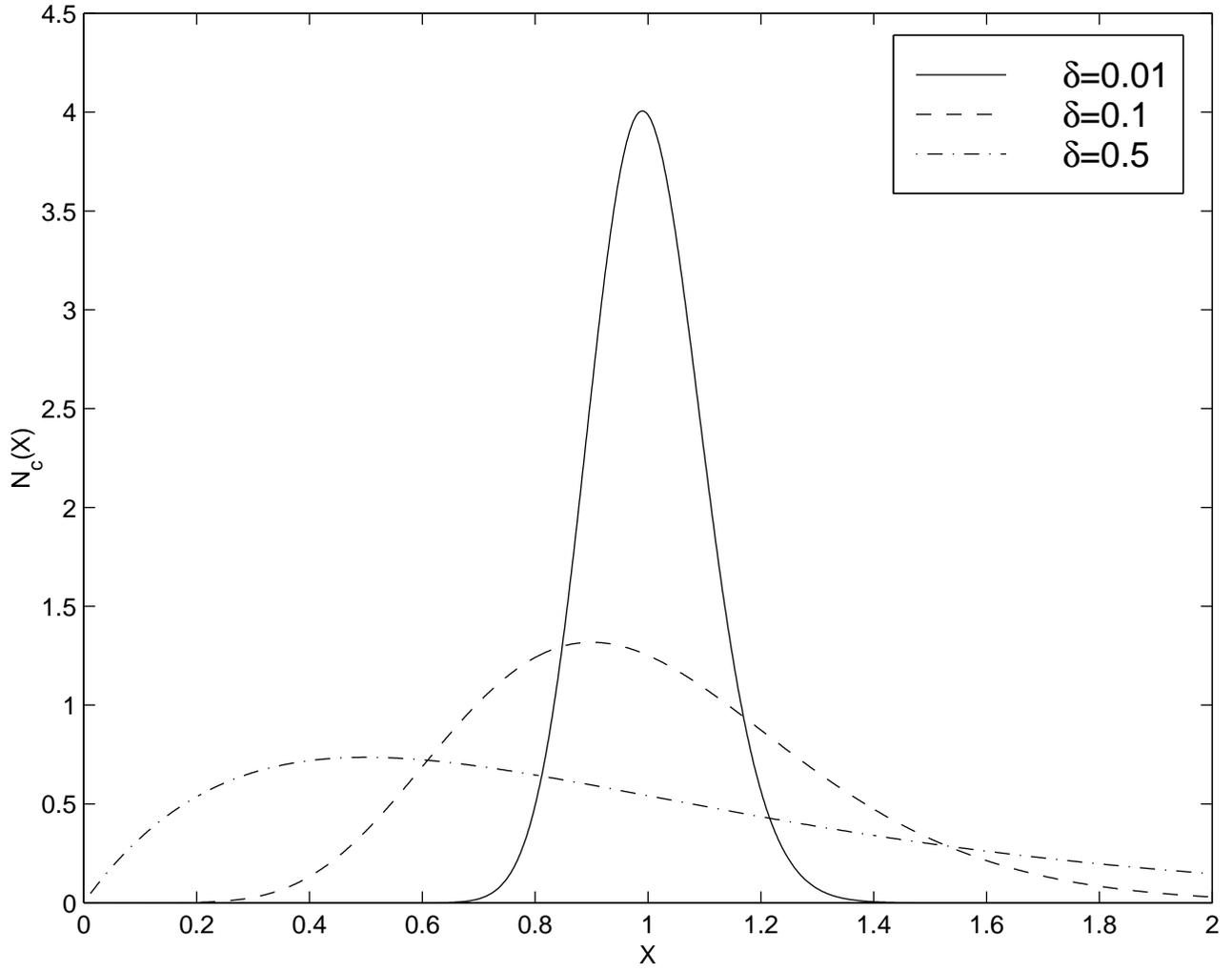}}
\caption{Steady state particle distributions for different turbulent
parameters}
\label{fig2}
\end{figure}
\begin{figure}
\centerline{\psfig{figure=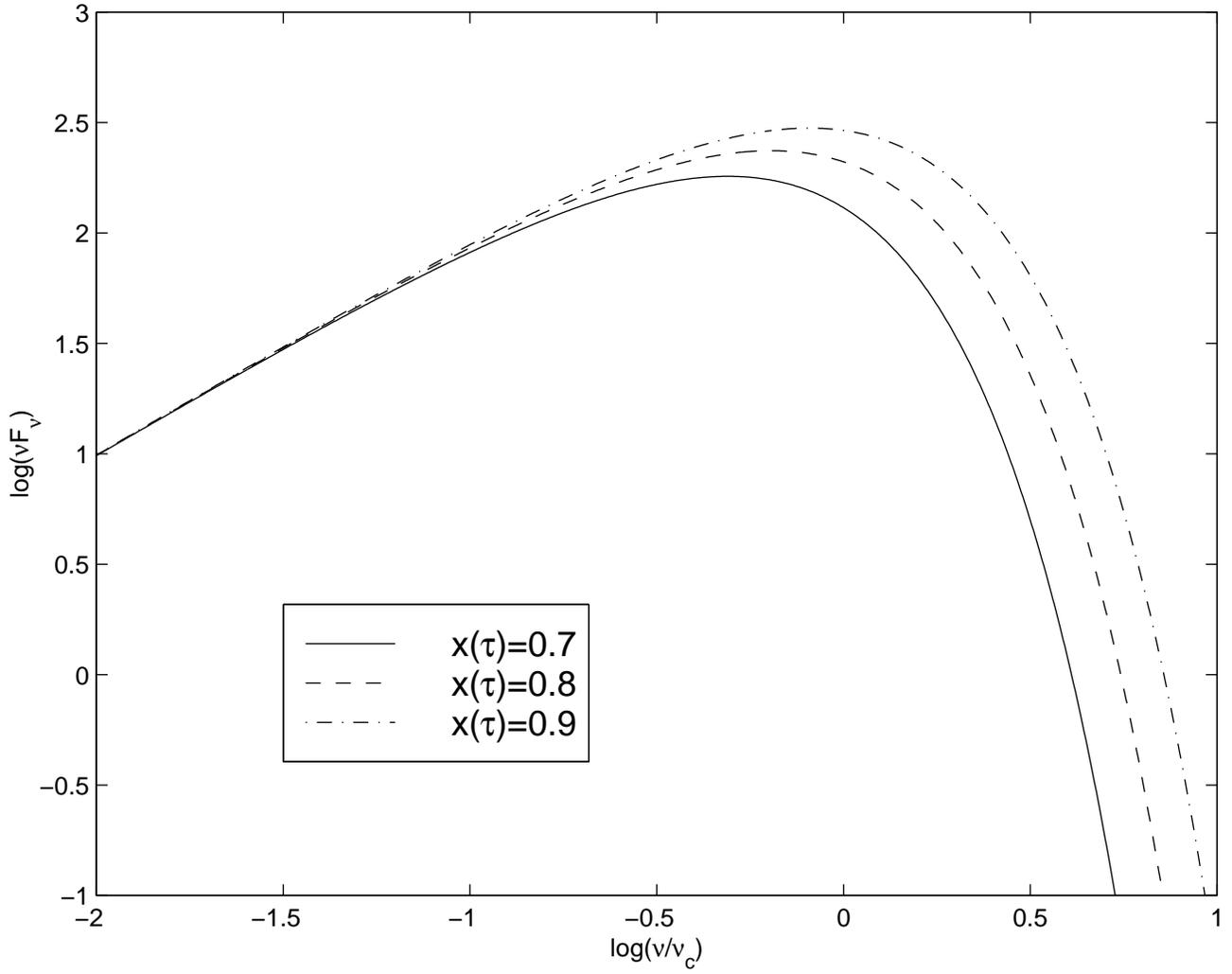}}
\caption{Synchrotron spectral evolution as the central energy of particle
distribution is evolving to higher energies}
\label{fig3}
\end{figure}
\begin{figure}
\centerline{\psfig{figure=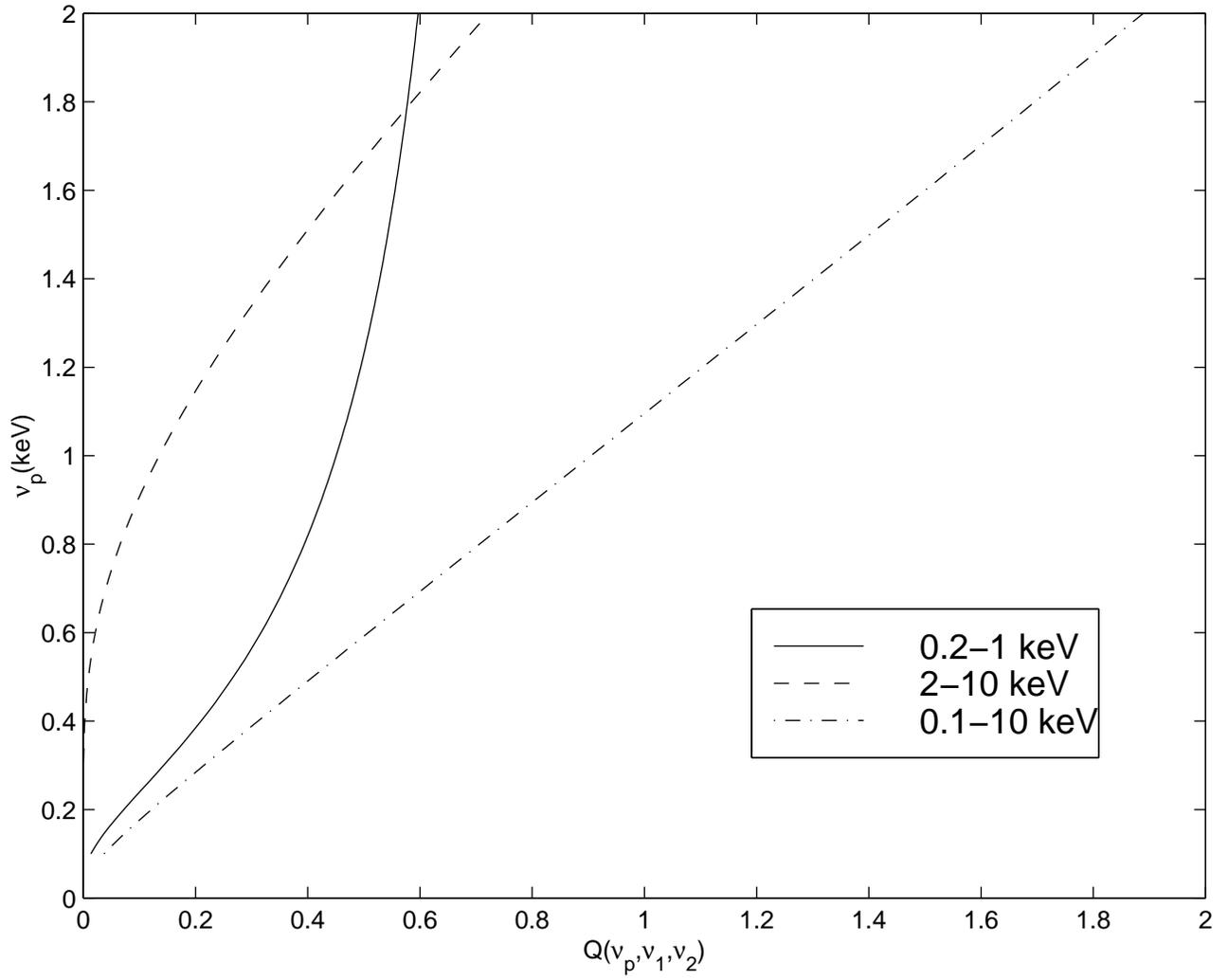}}
\caption{Synchrotron peak energy $\nu _p$ versus the fluxes $Q(\nu _p,\nu
_1,\nu _2)$ in specified frequency ranges $[\nu _1,\nu _2]$}
\label{fig4}
\end{figure}
\begin{figure}
\centerline{\psfig{figure=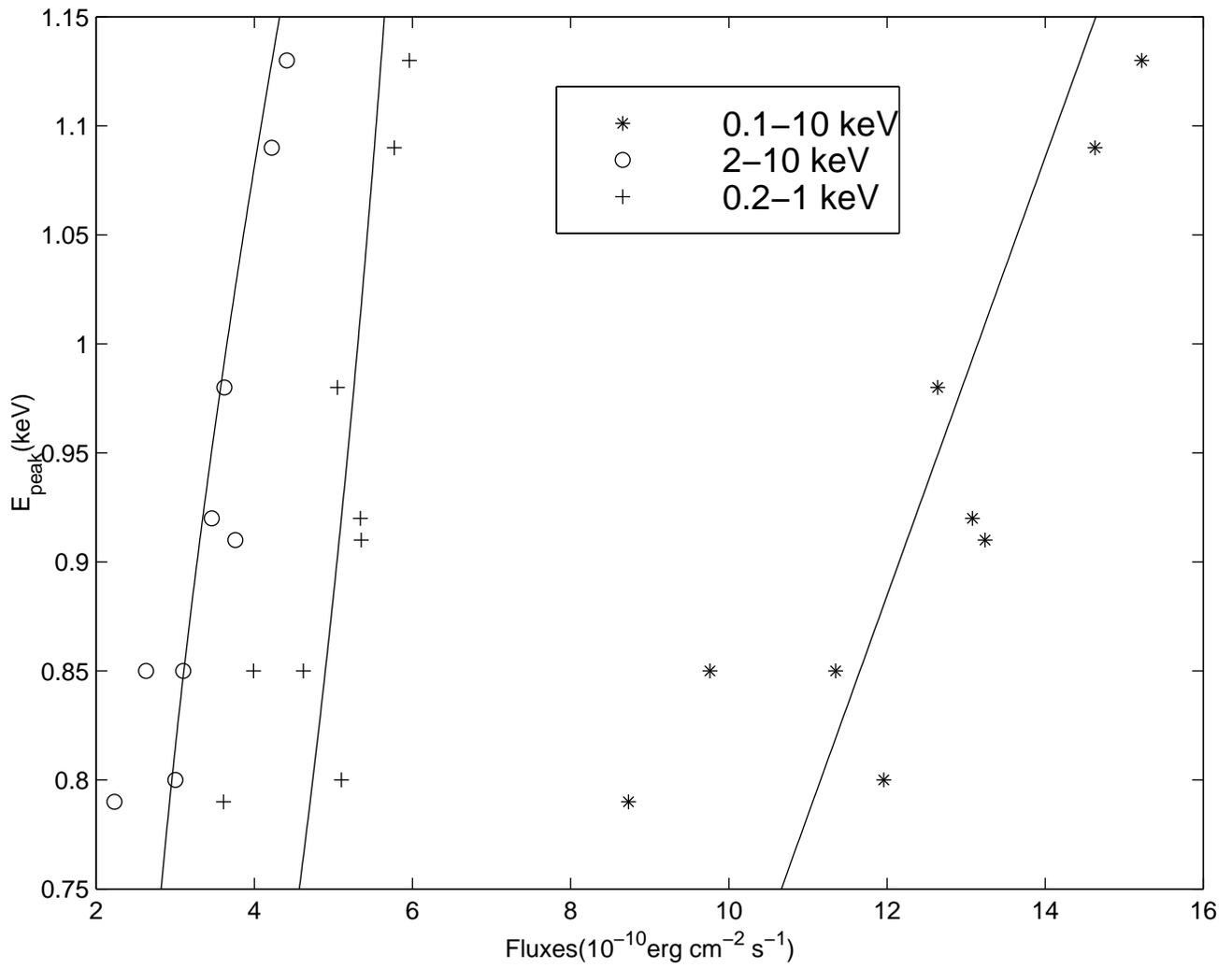}}
\caption{Synchrotron peak energy versus the fluxes
in specified frequency ranges for 1998 datasets of Mrk 421. The solid lines
represent the best fits of the model}
\label{fig5}
\end{figure}
\begin{figure}
\centerline{\psfig{figure=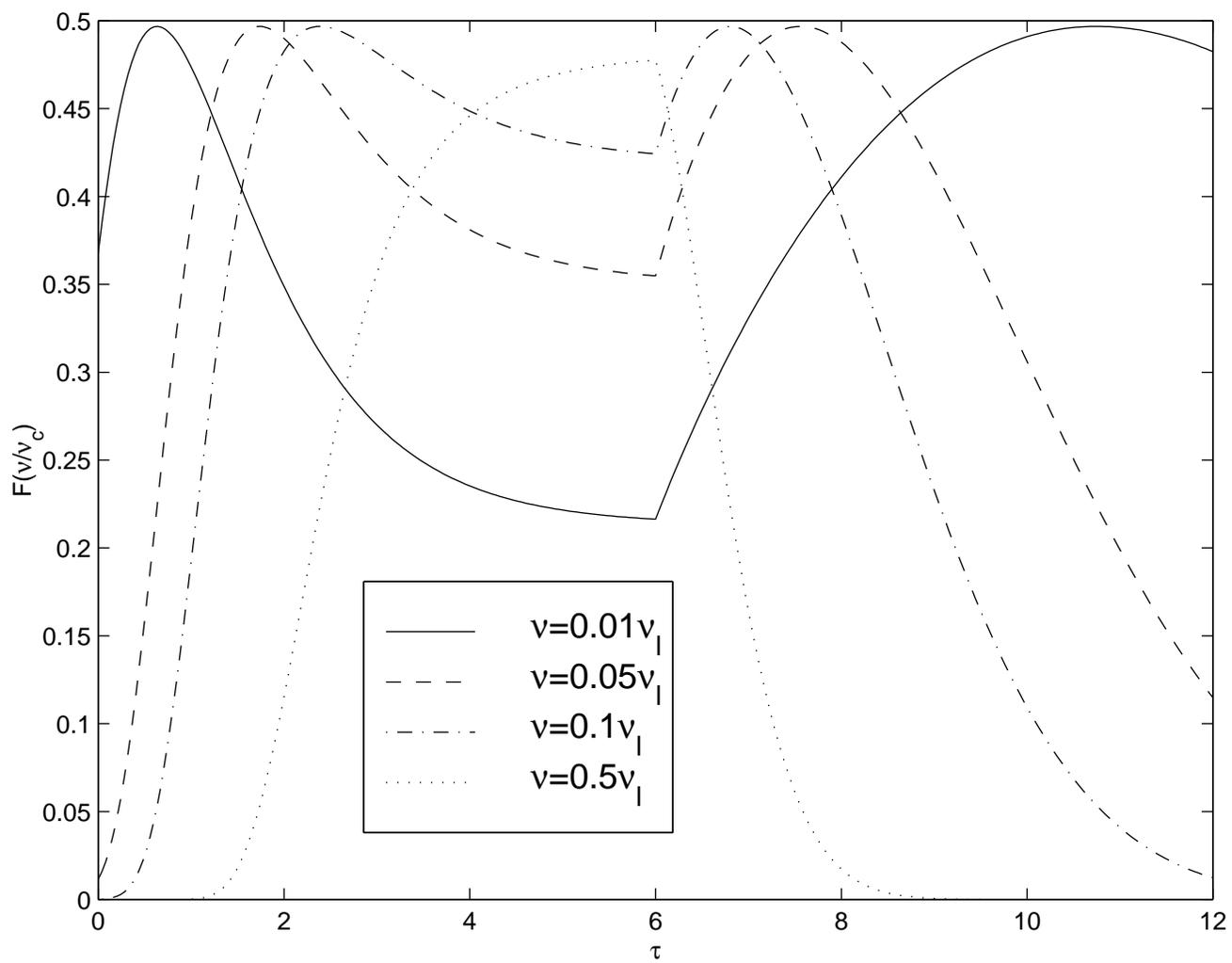}}
\caption{Multiwavelength light curves calculated from $F(\nu /\nu
_c)$ in the case of $X_0=0.1$ and $\tau _s=6.0$}
\label{fig6}
\end{figure}

\end{document}